11-15-2021

# E-Resource Management and Management Issues and Challenges

Subaveerapandiyan A
*Regional Institute of Education Mysore*, subaveerapandiyan@gmail.com

Ammaji Rajitha
*Central University of Tamil Nadu*, ammajirajitha810@gmail.com

Mohd Amin Dar
*Annamalai University*, mamin3787@gmail.com

Natarajan R
*Annamalai University*, cdmnatraj1972@gmail.com

# E-Resource Management and Management Issues and Challenges

Subaveerapandiyan A

Professional Assistant (Library)

Regional Institute of Education Mysore, India

E-mail: subaveerapandiyan@gmail.com

**Ammaji Rajitha**

Phd Research Scholar

Department of Library and Information Science

Central University of Tamil Nadu, India

E-Mail: ammajirajitha810@gmail.com

**Mohd Amin Dar**

PhD Research Scholar

Department of Library and Information Science

Annamalai University, India

E-mail: mamin3787@gmail.com

**R Natarajan**

Assistant Professor

Department of Library and Information Science

Annamalai University, India

E-mail: cdmnatraj1972@gmail.com

**Abstract**

E-resources are inevitable, technology has grown and libraries are also adopting the technologies although adopting have many challenges to the library professionals. Whenever something new comes they need to update themselves. A study investigated E-Resources management and management issues of Indian library professional perspectives. For this study, data was collected from various academic institutes/university libraries in India. It includes institutes of national importance, central, state, deemed and private universities. The study finds that the majority of the libraries subscribed to E-journals and E-books, administration related challenges faced by LIS professionals. The t-test results revealed a lack of professional skills is the reason for issues and challenges of Library management.

1. **Introduction**

E-Resources are essential in today's life of researchers, faculty members, research scholars, students, and information seekers. Information seeking online not only happens in a pandemic situation, but it also begins in the late 20th century. Day by day, user needs and demands are increasing gigantically. Facilitating the e-resources to the users, library professionals are facing various electronic resources management issues and challenges. Library users also sought the information online instead of searching the printed books they used to search e-resources. For that libraries are subscribing various e-resources and e-databases; library professionals are providing tremendous library e-services to the users such as content page service, current awareness service (CAS), electronic/virtual reference service, indexing and abstracting service, E-library, E-journals, E-database, E-bibliographic database, E-book, E-mail, Web OPAC and so on. Libraries develop various services to fulfil user needs so that library employees face technical, administrative, and management issues. ERM changes library professionals' work from traditional to digital because many institutes face various problems such as lack of staff, skill and budget. Development of technologies user demands also depends on more E-resource services. The digital environment divides the attachment of print and electronic.

2. **Earlier Studies**

Morales & Beis (2021) conducted a survey study on common communication electronic resources issues. In their study academic librarians are reported mostly faced with acquisition of e-resources, evaluation and renewal was an issue of the library professionals. They give suggestions to improve in some areas including policy and procedure, evaluation of e-resources, need to develop structure models and facilitate communication between libraries and users, librarians and publishers, collaborative projects. Finally, they conclude lack of knowledge of professionals and more training is required.

Abdulla & Devi (2020) discussed in their research paper on E-resources management of selective universities in Kerala. They did a literature survey on five Kerala universities. Their study analysed e-resources services, ICT facilities, subscribed e-resources, tools used for selecting e-resources, general criteria followed by selecting the e-resources, technical aspects of e-resources, and e-resources management procedures, evaluating aspects and interface for accessing e-resources. The study finds out that they subscribed to all five universities to e-journals and E-databases, and the OPAC facility provides all the universities. Commonly, a publisher catalogue is used for the selection of e-resources and trial offers based. Other criteria for choosing e-resources are foremost user needs and Literature availability format. Technical perspectives e-resources library preferred multi-user facility of access the resources, IP based access, software and hardware compatibility, viewing facility of usage stat. Interfaces of accessing e-resources most of the library subscribed to DOAJ, SpringerLink, J-Gate Plus and Wiley Online Library.

Lo et al. (2017) conducted a study on progressive development ERM and used a qualitative interview method for their study. They discussed issues, problems, good practices in Hong Kong Libraries and required technical skills. Interview revealed e-resources saves time, space, manpower and are error-free. Library faced challenges they are subscription cost is more and budget is a problematic one and also still many of the user's demands print books so finally, the result show user needs both print as well as digital because of this library faces financial related issues.

Sohail & Ahmad (2017) did a survey study on awareness of electronic resources used by faculties and students in Fiji National University. In their study, they found the students and faculties met the problem was poor IT infrastructure and website is not supporting properly because of troubleshooting issues and heavy traffic of E-resources access. Additionally, they found out and pointed out lack of library professional's skills is also the reason for inadequate library services.

3. **Materials and Methods**

The study was conducted with Indian library professionals for that stratified sampling method used, and a structured questionnaire was prepared and distributed with the help of Google form through email. The questionnaire comprises three parts. Part-I: Socio-Demographic Details, Part-II: E-Resources and Library services, Part-III: E-Resource Challenges and Management Issues. For this study, we received 156 responses from various academic institutes/university libraries. They are Institutes of National Importance such as IIT, AIIMS, IIM, NIT, IIIT, IISER, NIPER, NID, SPA, five central universities and four medical research institutes, Central Universities, State Universities, Private Universities, Deemed Universities. For measurement of the study, 5 points, the Likert Scale was used, 1 Strongly Disagree, and 5 Strongly Agree. For data analysis purposes, we used SPSS software.

4. Data Analysis and Interpretation

Table 1. Gender

| Gender | Respondents | Percentage |
|---|---|---|
| Female | 44 | 28.2 |
| Male | 112 | 71.8 |

Table 1 shows the gender-wise of the 156 respondents. 28.2% (44) were female and 71.8% (112) were male.

Table 2. Highest Educational Qualifications

| Academic Qualification | Respondents | Percentage |
|---|---|---|
| Graduated | 2 | 1.3 |
| Post Graduated | 98 | 62.8 |
| M.Phil. | 12 | 7.7 |
| PhD | 44 | 28.2 |

Table 2 describes the academic educational qualification of the respondents. Results revealed that most of them post-graduated 62.8%.

Table 3. Working Institute Type

| Institute types | Respondents | Percentage |
|---|---|---|
| Institute of National Importance | 32 | 20.5 |
| Central Universities | 28 | 18 |
| State Universities | 23 | 14.8 |
| Deemed Universities | 26 | 16.6 |
| Private Universities | 47 | 30.1 |

Above table 3 provides the details of respondents working educational institute types. Most of the respondents were from Private Universities 30.1% (47), 20.5% (32) respondents belong to the Institute of National importance. 18% (28) of respondents were from central universities, 16.6% (26) of respondents from deemed universities, 14.8% (23) of respondents from state universities.

**Table 4. LIS Professional Present Designation**

| Job Designation | Respondents | Percentage |
|---|---|---|
| University Librarian/Head Librarian/Library Manager/Director | 31 | 19.9 |
| Deputy Librarian/Associate Director | 10 | 6.4 |
| Assistant Librarian/Senior Assistant Librarian | 58 | 37.2 |
| Professional Assistant/ Semi Professional Assistant/ Library Assistant | 45 | 28.8 |
| Library Technician | 2 | 1.3 |
| Others | 10 | 6.4 |

Above table 4 describes respondents' designation in their working institutes. 37.2% of respondents are Assistant Librarian/Senior Assistant Librarian rank, 28.8% Professional Assistant/ Semi Professional Assistant/Library Assistant, 19.9% University Librarian/Head Librarian/Library Manager/Director, 6.4% Deputy Librarian/Associate Director and Others include Library Trainee and Library Attenders, 1.3% Library Technician.

**Table 5. Library Work Experience**

| Library Work Experience | Respondents | Percentage |
|---|---|---|
| Less than 1 year | 3 | 1.9 |
| 1-3 years | 8 | 5.1 |
| 4-5 years | 15 | 9.6 |
| 6-10 years | 35 | 22.4 |
| More than 10 years | 95 | 60.9 |

Table 5 inferred respondents' library working experience in years. The majority of the respondents had 60.9% over 10 years of experience, 22.4% 6 to 10 years of experience, 9.6% 4 to 5 years of experience, 5.1% 1 to 3 years of experience and 1.9% of respondents less than one-year experience in the library.

**Table 6. Does your Library Provides E-resources Services to the users?**

| Library Provides E-resource services | Respondents | Percentage |
|---|---|---|
| Yes | 156 | 100 |
| No | 0 | 0 |

We found it from the above table 6 library provides e-resources services to the users. The majority of the respondents 100% say 'yes' and 0% say 'no'.

**Table 7. Does your Institute subscribe below-mentioned multidisciplinary databases?**

| Does your Institute subscribe below-mentioned multidisciplinary databases? | Yes | No |
|---|---|---|
| Scopus | 76 (48.7%) | 80 (51.3%) |
| Web of Science | 80 (51.3%) | 76 (48.7%) |
| ScienceDirect | 96 (61.5%) | 60 (38.5%) |
| JSTOR | 110 (70.5%) | 46 (29.5%) |
| ProQuest Central | 69 (44.2%) | 87 (55.8%) |
| EBSCOhost | 88 (56.4%) | 68 (43.6%) |
| Gale | 24 (15.4%) | 132 (84.6%) |

**Figure 1: Multidisciplinary database subscribed by the libraries**

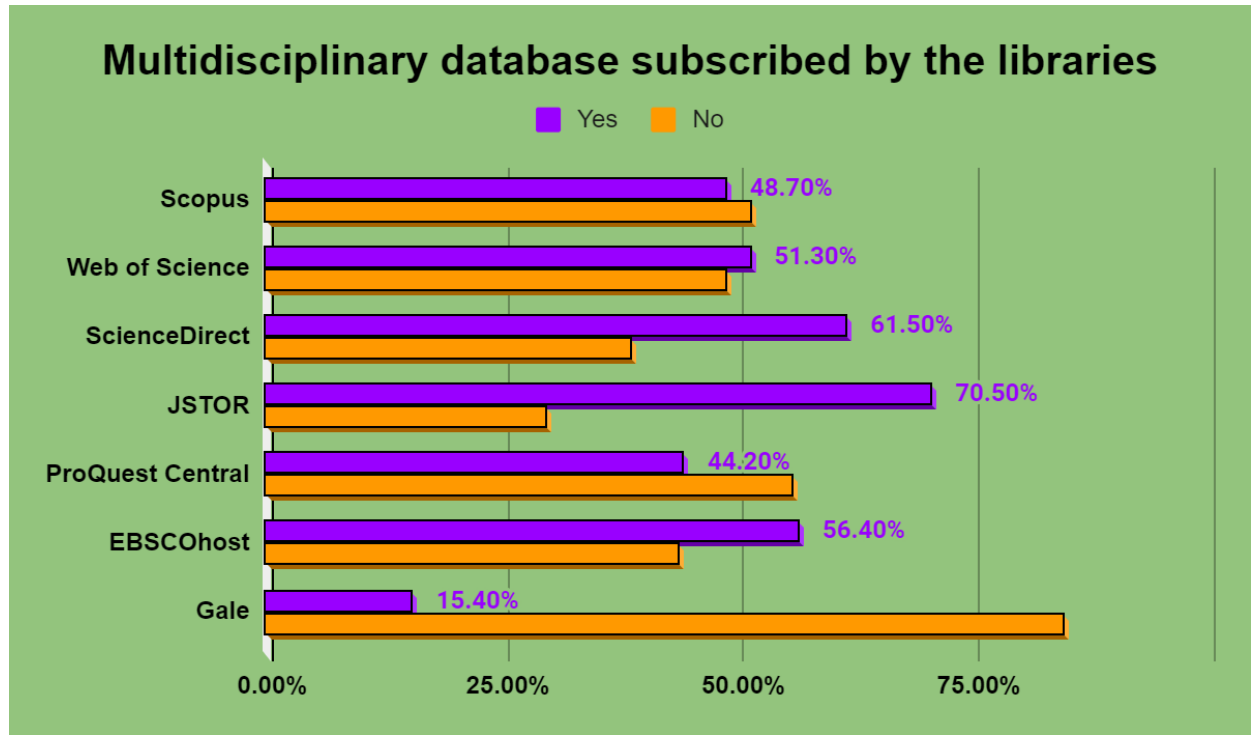

Above table 7 and figure 1 shows that libraries subscribed to multidisciplinary databases. In this majority of the library subscribed JSTOR 70.5% and least subscribed is Gale 15.4%.

**Table 8. User electronic services provided by your library**

| User electronic services provided by your library | Respondents | Percentage (n=156) |
|---|---|---|
| Bulletin-Board service | 67 | 42.9 |
| Content page service | 79 | 50.6 |
| Current awareness service (CAS) | 122 | 78.2 |
| Electronic/virtual reference service | 115 | 73.7 |
| Indexing and abstracting service | 71 | 45.5 |
| Internet-enabled workstations | 96 | 61.5 |
| Intranet and Internet service | 123 | 78.8 |
| Other bibliographical services and on-demand | 105 | 67.3 |

| Selective dissemination of information (SDI) | 99 | 63.5 |
| Training programme on information literacy | 110 | 70.5 |

**Figure 2: User electronic services provided by your library**

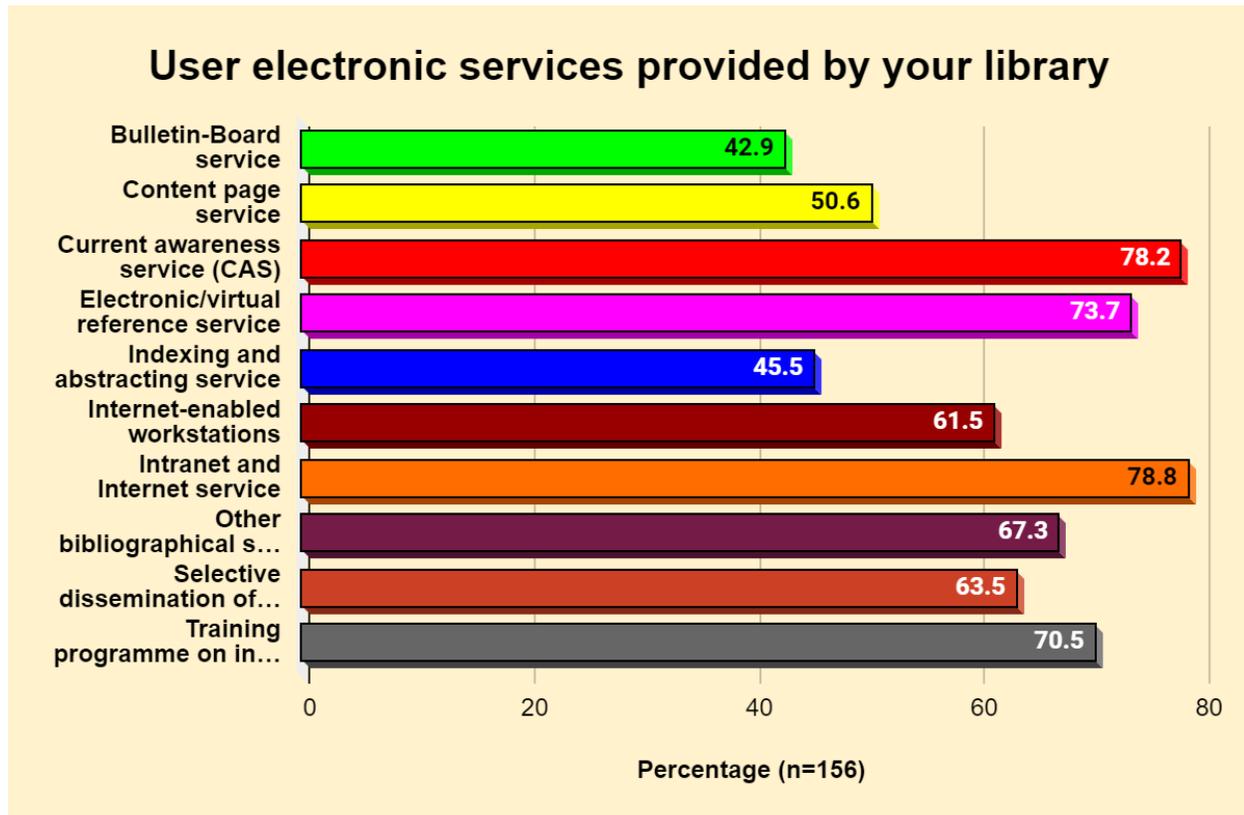

Table 8 and figure 2 describes electronic user services provided by their institute library. Respondents were asked to choose the services, and ten options were given. The study present that 78.8% of libraries provides internet and intranet services, followed by 78.2% current awareness services, 73.7% electronic and virtual reference services, 70.5% Training programme on information literacy, 67.3% Other bibliographical service and on-demand, 63.5% Selective dissemination of information (SDI), 61.5% Internet-enabled workstations, 50.6% Content page service, 45.5% Indexing and abstracting service, and finally the least respondents 42.9% Bulletin-Board service

**Table 9. E-resources user services**

| E-resources user services | Respondents | Percentage (n=156) |
|---|---|---|
| E-library | 116 | 74.4 |
| E-book | 146 | 93.6 |
| E-clipping | 74 | 47.4 |
| E-data archives | 70 | 44.9 |
| E-database | 133 | 85.3 |
| E-dictionary | 57 | 36.5 |
| E-images | 51 | 32.7 |
| E-journals | 147 | 94.2 |
| E-mails | 126 | 80.8 |
| E-patents | 33 | 21.2 |
| E-quick reference | 63 | 40.4 |
| E-standards | 40 | 25.6 |
| E-theses | 110 | 70.5 |
| E-zines | 37 | 23.7 |

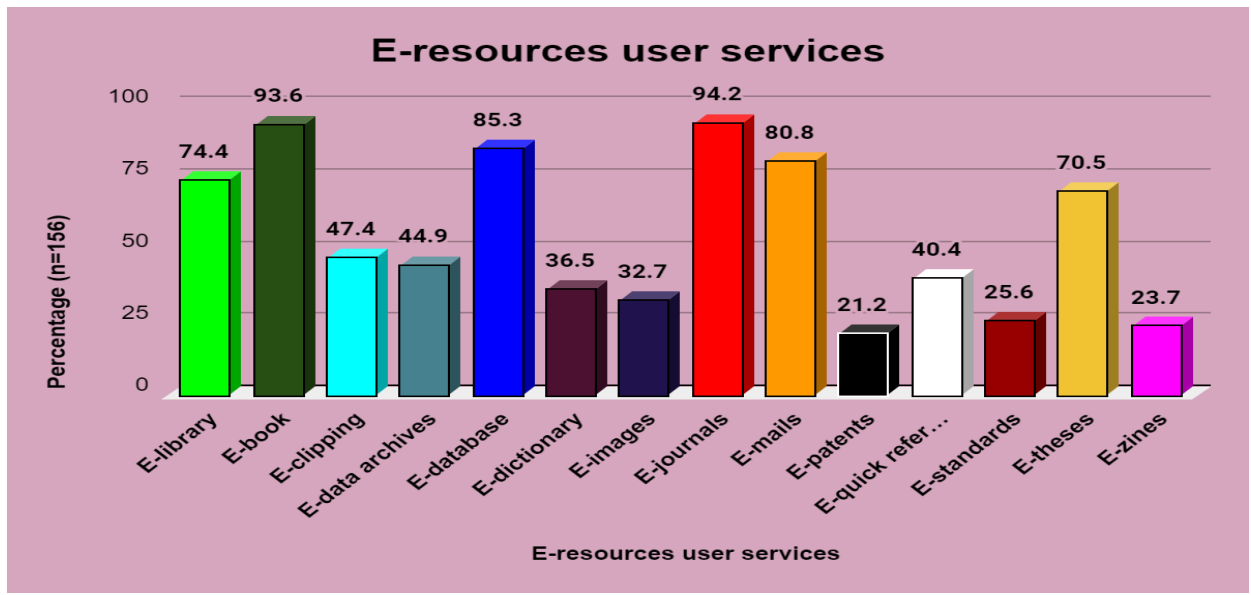

**Figure 3: E-resources user services provided by the library**

The above table 9 and figure 3 show, respondents were asked to mention libraries providing E-resources services to the users. Options were given and free to choose from 13. Over 90% of respondents chose E-journals and E-Book services, over 80% of respondents chose E-database and Email services, over 70% of respondents chose E-library and E-theses. The remaining services are less than 50% of libraries only offering.

**Table 10. What other general types of workload are you faced with?**

| Workload faced by Library Professionals | Respondents | Percentage (n=156) |
|---|---|---|
| Administration | 108 | 69.2 |
| Assessment/user experience | 77 | 49.4 |
| Cataloging/metadata creation | 87 | 55.8 |
| Circulation/ILL | 83 | 53.2 |
| Collection development/management | 106 | 67.9 |
| Digitization | 98 | 62.8 |
| Institutional Repository Management | 84 | 53.8 |
| Instruction | 48 | 30.8 |
| Print acquisitions | 74 | 47.4 |
| Print serials management | 60 | 38.5 |
| Reference | 89 | 57.1 |
| Web services (library website) | 88 | 56.4 |

**Figure 4: Workload faced by Library Professionals**

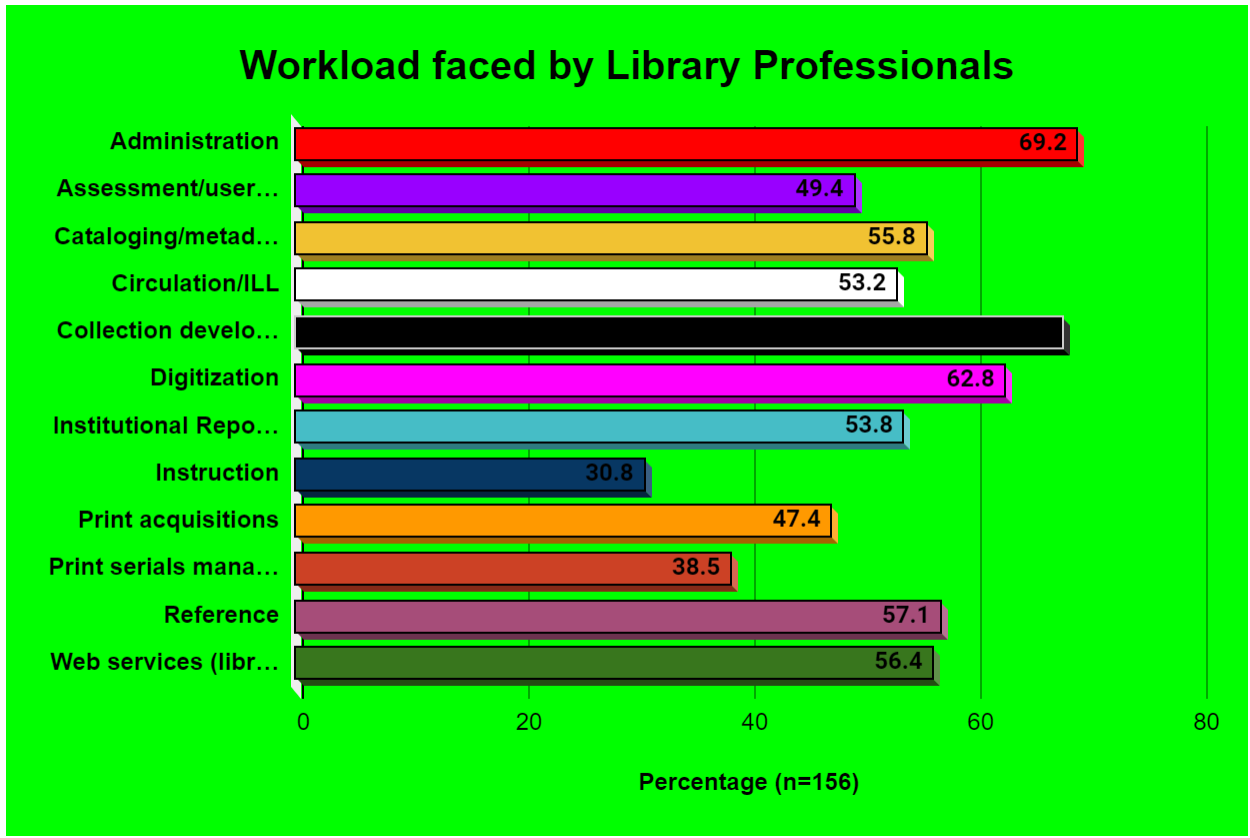

Table 10 and figure 4 describes the workload faced by Library Professionals. The highest number of respondents 69.2% faced administration, 67.9% Collection development/management, 62.8% Digitization, 57.1% reference, 56.4% Web services (library website), 55.8% Cataloging/metadata creation, 53.8% Institutional Repository Management, 53.2% Circulation/Interlibrary loan, 49.4% Assessment/user experience, 47.4% Print acquisitions, 38.5% Print serials management, 30.8% Instruction.

**Table 11. Challenges of digital resources management work experience**

| Challenges of digital resources management | Mean | SD | T-Value |
|---|---|---|---|
| Budget planning for e-resources acquisition | 3.32 | 0.75 | 6.38 |
| Vendor Management | 2.96 | 0.85 | 9.12 |
| Publisher management | 2.89 | 0.90 | 9.48 |
| Managing journal packages | 3.11 | 0.82 | 7.96 |
| Tracking gaps in journals | 2.98 | 0.97 | 8.58 |
| Renewal of E-Resources | 3.06 | 0.97 | 7.93 |
| Price negotiation | 3.26 | 0.81 | 6.78 |
| Shortage of staff | 3.08 | 1.07 | 7.48 |
| Lack of professional skills | 2.72 | 1.22 | 9.59 |
| Troubleshoot management | 2.95 | 0.89 | 9.04 |

From Table -11, it is inferred that the biggest challenge that the people with experience faced is the lack of professional skills followed by troubleshooting management. Professional skills are the huddles for professionalism in managing digital resources. This means that even the well-experienced professionals and entrants face problems related to their professional skills. The highest mean value 3.32, signifies that Budget planning for e-resources acquisition is the challenge most people face. However, the standard deviation is marginal with price negotiation.

**Table 12. Challenges of digital resources management educational qualification**

| Challenges of digital resources management | Mean | SD | T-Value |
|---|---|---|---|
| Budget planning for e-resources acquisition | 3.32 | 0.75 | -10.09 |
| Vendor Management | 2.96 | 0.85 | -5.89 |
| Publisher management | 2.89 | 0.90 | -5.12 |
| Managing journal packages | 3.11 | 0.82 | -7.55 |
| Tracking gaps in journals | 2.98 | 0.97 | -5.71 |
| Renewal of E-Resources | 3.06 | 0.97 | -6.48 |
| Price negotiation | 3.26 | 0.81 | -9.12 |
| Shortage of staff | 3.08 | 1.07 | -6.3 |
| Lack of professional skills | 2.72 | 1.22 | -2.88 |
| Troubleshoot management | 2.95 | 0.89 | -5.7 |

From Table -12, it is evident that e-resource acquisition and price negotiations are the challenges most people face. Professional skill is not the problem with this group. However, the pricing is the major. Determinant. But the standard deviation is of utmost uniform at all the challenges

**5. Findings of the Study**

- The present study enlightens that 71.8% of the highest number of respondents are males.
- It determined that the highest 62.8% of respondents have Post Graduation as higher education in all institutions.
- The study showed that 30.1% of LIS professionals work in private universities.
- It enlightens that 37.2% of respondents have the designation of assistant librarian/senior assistant librarian.
- The study found that 60.9% of respondents have a minimum ten years of work experience in their field.
- The study result showed that 100% of libraries are providing E-resource services to their users.
- The study found that 70.5% of libraries subscribed to JSTOR, and the least 15.4% subscribed to Gale.
- It enlightens that 78.8% of libraries provided Internet and Intranet services to their users.
- The study showed that 90% of respondents preferred E-resources services to access their reading materials.
- The study found that 69.2% of the respondents face the problem of the workload from the administration.
- The t-test study result revealed that Lack of professional skills and troubleshooting management.
- The t test result shows e-resources acquisitions and price negotiations are the major challenges.

## 6. Conclusion

In this research, ten challenges for librarians and staff of the library in managing digital resources are considered. The study highlights lack of professional skills and publisher management on digital resources is the biggest challenge for professionals with high work experience and required qualifications. While the digital technologies areal to library management, strategies focused on the renewal of e-resources. Improved vendor management, publisher management, and troubleshoot management are required for libraries (or library staff) aspiring to digital transformations. This analysis codifies an integrated approach for library management, including library staff management, compliances, digital operations and reports. A deep understanding of technology with adequate staff is essential for this kind of transformation.